\documentclass[aps,preprint,showpacs,groupedaddress]{revtex4}  % for double-spaced preprint
\usepackage{graphicx}  % needed for figures
\usepackage{dcolumn}   % needed for some tables
\usepackage{bm}        % for math
\usepackage{amssymb,amsfonts,amsmath}

\newcommand{\ave}[1]{\langle #1\rangle}
\newcommand{\eq}[1]{Eq.\,(\ref{#1})}

\newcommand{\der}[1]{\partial_{#1}}
\newcommand{\PS}{{\cal P}}

\newcommand{\strong}{\gamma\!\rightarrow\!\infty}

\begin{document}

% the following line is for submission, including submission to the arXiv!!
%\hspace{5.2in} \mbox{Fermilab-Pub-04/xxx-E}

\title{Multi-species stochastic models: \\small RNA regulation of gene expression}
\date{\today}

\begin{abstract}
The processes, resulting in the transcription of RNA, are intrinsically noisy. It was observed experimentally that the synthesis of mRNA molecules is driven by short, burst-like, events. An accurate prediction of the protein level often requires one to take these fluctuations into account. Here, we consider the stochastic model of gene expression regulated by small RNAs. Small RNA post-transcriptional regulation is achieved by base-pairing with mRNA. We show that in a strong pairing limit the mRNA steady state distribution can be determined exactly. The obtained analytical results reveal large discrepancies in the deterministic description, due to transcriptional bursting. The proposed approach is expected to work for a large variety of multi-species models.
\end{abstract}

\pacs{}
\maketitle 

%\section{\label{sec:level1}First-level heading}
% sections are not used for PRL papers

Models describing nonlinear interactions between several species of particles are widely spread in many areas of science. Let us mention a couple of the famous examples: the Predator-Prey (PP) model in biology and the Susceptible-Infected-Removed (SIR) epidemic model. Quantitative description of gene regulation by small RNA also requires the analysis of a nonlinear two-species interacting system.

Small noncoding RNAs (sRNAs) are generally synthesized as independent transcripts that interact with mRNA. The functional role of many of those sRNAs is the inhibition of translation by base pairing with the mRNA in the ribosome-binding site, for review see \cite{gottesman05,waters09}. Hence, the sRNA and mRNA two-species recombination process has to be integrated in the gene expression model.

The models mentioned above are inherently random. Stochasticity usually arises due to a low abundance of the interacting species, since they are present in {\em discreet} numbers. As a result, the deterministic models often fail to correctly predict the characteristics of the system for a certain range of parameters. 

%In order to account for noise due to the discreetness of the particles, the probabilistic Master equation is frequently employed. A Master equation describes the evolution of the probability distribution to find any given particle number $n_{_X}$ of each species $X$ in the system. The dynamical rules of interactions are interpreted now as probabilistic processes.

In order to account for noise due to the discreetness of the particles, the dynamical rules of evolution are interpreted as probabilistic processes.
In many instances, transition probabilities, governed by these dynamical rules, are nonlinear in terms of the number of particles involved.

However, in the modeling of the multi-species systems the only source of nonlinearity is often the interaction between {\em different} species. Interactions within the same species are usually modeled by the linear birth-death processes which are sufficient to capture the system's dynamics. In order to model inter-species interaction, one is bound to introduce nonlinear kinetic terms.

Even single site models, that describe nonlinear, interacting multi-species particles, are difficult to treat analytically. An exact solution can be usually derived only for models with special symmetries (conservation laws).

If the symmetries between different species are broken by either dynamical rules or the diffusion for spatially extended problems, then there is little hope to achieve an exact solution for a multi-species  nonlinear model.

Nevertheless, several powerful approximation schemes were developed to treat various aspects of multi-species problems analytically. They usually rely upon the separation of time-scales in the model \cite{samoilov05,kamenev08,parker09}. This separation sometimes allows one to average out one random variable and reduce the system into an effectively single species problem. The reduced problem, however, usually remains nonlinear and further approximations may be required in order to solve it.

We propose a similar approach based on the time-scale separation, that leads to a {\em linear}, coupled multi-species model. Based on several readily solvable examples, we believe that our method can be proved useful in dealing with a large variety of multi-species systems. The main advantages of the proposed method are symmetry preservation and linearity of the effective evolution equations.

In many biological and epidemic models, the coupling $\gamma$ controlling the interaction strength is large in comparison with the rest of the kinetic parameters in the system. For example, the capture of prey by the predator happens on a much shorter time-scale than the birth and death of both species.

Exactly the same argument is applicable to some epidemic models. The typical contraction time is fairly short, compared to the time-scale of the recover process. Similarly, in the model of gene regulation by sRNA recombination rate is often large.

Surprisingly, strong interaction may lead to the simplification of the model! Indeed, in the limit of the infinitely large interacting coupling $\strong$, only {\em one} species may be present at a time. For the PP model, this means that at any site of the lattice either predators or prey are present, not both at the same time, since predation is instantaneous.

Since in many models the only source of nonlinearity is the inter-species interaction term, the limit $\strong$ results in set of linear coupled evolution equations for each species. Moreover, this limit preserves all the symmetries of the model at hand. Hence, even for models with weak inter-species interaction, the obtained equations can be used to identify, say, critical exponents of the system.

To demonstrate how an exact solution may be achieved in the strong interaction limit, we consider the very interesting problem of post-translational gene regulation. Unlike PP and SIR models, noninteracting terms can be, in this case, rather complicated due to transcriptional bursting. Nevertheless, the sRNA regulation model can be solved exactly in the strong interaction limit.

\begin{figure}
\includegraphics[scale=0.3]{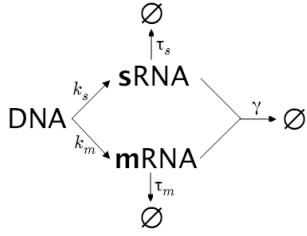}
\caption{The model of sRNA regulation. Both mRNA and sRNA molecules are synthesized independently with arrival rates $k_m$ and $k_s$ correspondingly. They decay either naturally or as a result of mutual degradation with rate $\gamma$.}\label{fig:srna}
\end{figure}

The model of sRNA regulation is depicted in Fig.~\ref{fig:srna} and the {\em deterministic} (mean-field) evolution equations for the corresponding RNA levels are \cite{levine07}
\begin{align}
	\der{t}m = k_m - \tau_m^{-1}m - \gamma m s, \label{mf-m}\\
	\der{t}s = k_s - \tau_s^{-1}s - \gamma m s.
\end{align}
Here $k_j$ and $\tau_j^{-1}$ are corresponding arrival and decay rates for species $j=\{m,s\}$.

In terms of the Master equation, these rates can be interpreted as probabilities per unit time of the synthesis and decay events.

A pair of molecules of mRNA and sRNA can recombine into an inert complex that quickly decays with the probability rate $\gamma$ \cite{levine07}. In what follows, we assume that the rate $\gamma$ is fast compared to the synthesis and natural decay rates of both species mRNA and sRNA.

If the rate $\gamma$ is infinite, it is clear that, at any given time, at most one species of particles is present. Therefore, a probability to find $m$ mRNA and $n$ sRNA molecules at any time $t$ is the sum
\begin{align}
	\PS(m,n; t) = \delta(n)\PS_m(m; t) + \delta(m)\PS_s(n; t),\label{master}
\end{align}
where $\delta(n)$ is the Kronecker delta function, which has the value $1$ for $n=0$, and zero otherwise. Here, $\PS_{m}$ and $\PS_{s}$ are the reduced probability distribution for mRNA and sRNA molecules.

Let us now derive a detailed balance condition for the reduced mRNA probability distribution, which is essentially a Master equation at the steady state. To this end, we assume that RNA synthesis for both species occurs in bursts. 

Indeed, in recent exciting experiments, several groups were able to probe the statistics of various cellular components for a {\em single} DNA molecule \cite{golding05,cai06,yu06}. 
It was observed in many instances that the synthesis of mRNA molecules is driven by short, burst-like, events \cite{golding05}.

For $n\ge 1$, the detailed balance condition reads 
\begin{align}
	k_m \sum_{i=1}^n \rho_m(i)\PS_m(n-i) + k_m 
	\sum_{i=n}^\infty \rho_m(i)\PS_s(i-n)\nonumber\\
	+\tau_m^{-1}(n+1)\PS_m(n+1) + k_s\sum_{i=1}^\infty \rho_s(i)\PS_m(n+i) \nonumber\\
	= \left[k_m + k_s + \tau_m^{-1}n\right]\PS_m(n).\label{master-m}
\end{align}
Here, $\rho_m(i)$ is the conditional probability distribution to find $i$ mRNA molecules in the burst, provided that the burst is not empty. The analogous quantity, $\rho_s(i)$, characterizes the sRNA burst distribution.

The mRNA detailed balance condition \eq{master-m}, and its symmetric counterpart for sRNA, fully describe the steady state distribution of the system, once the burst statistics  $\rho(n)$ are given for each species.

In several experiments, various microscopic parameters of the gene expression model were obtained, such as, the mean burst frequency and the mean number of mRNAs in the burst (transcript number) \cite{golding05,cai06}.
In bacteria, the number of mRNA molecules detected in synthesis bursts was governed by a geometric probability distribution \cite{golding05}
\begin{align}
	\rho_m(n) = (1-u_m)\, u_m^{n-1},\quad n=1,2,...\label{cond-geometric}
\end{align}

It is, then, reasonable to expect that sRNA burst statistics $\rho_s$ is also given by the geometric distribution with a different control parameter, $u_s$.

In order to solve \eq{master-m}, it is convenient to define burst and steady state generating functions
\begin{align}
	g_j(x) \equiv \sum_{n=0}^\infty x^n \rho_j(n) = \frac{(1-u_j)x}{1-u_j x},\label{gen-burst}\\
	G_j(x) \equiv \sum_{n=0}^\infty x^n \PS_j(n),\label{gen}
\end{align}
for each species $j=\{m,s\}$. Note that, due to the definition \eq{gen}, the generating functions are analytic in the interval $x\in [-1,1]$, since $0\le \PS_j(n) \le 1$ for any integer $n$.

The average quantities can now be derived by evaluating the derivatives of these generating functions at point $x=1$, e.g.,
\begin{align}
	n_m \equiv \sum_{n=1}^\infty n\rho_m(n) = \der{x}g_m(1) = \frac{1}{1-u_m},\\
	\ave{m}_r \equiv \sum_{n=0}^\infty n \PS_m(n) = \der{x}G_m(1),
\end{align}
where $n_m$ and $\ave{m}_r$ are the mRNA transcript number and regulated mean level correspondingly.

For geometric burst distributions we derive from \eq{master-m}
\begin{align}
	\der{x}G_m(x) = \tau_m k_m \frac{G_m(x)+G_s(u_m)}{1-u_m x} -\nonumber\\
	\tau_m k_s \frac{G_m(x) - G_m(u_s)}{x-u_s}.\label{steady-m}
\end{align}
The symmetric equation, with interchanged indices $m$ and $s$, holds for the sRNA steady state generating function.

The point $x=u_s$ corresponds to blow-up in the {\em normalized} homogeneous solution of \eq{steady-m}
\begin{align}
	h_m(x) \equiv \left(\frac{1-u_s}{x-u_s}\right)^{k_s \tau_m}\left(\frac{1-u_m}{1-u_m x}\right)^{\frac{k_m \tau_m}{u_m}}.
\end{align}
The analyticity of functions $G_m(x)$ and $G_s(x)$ on the entire interval $\{-1,1\}$ means that the following condition should be satisfied in order to remove this singularity:
\begin{align}
	G_m(1) = I_m^1(u_s,1) G_m(u_s) + I_m^2(u_s,1) G_s(u_m).\label{analyticity}
\end{align}
Here we defined the finite integrals for particular solution representation:
\begin{align}
	I_m^1(a,b) \equiv \int_a^b\mathrm{d}x\, \frac{k_s\tau_m}{(x-u_s)h_m(x)},\\
	I_m^2(a,b) \equiv \int_a^b\mathrm{d}x\, \frac{k_m\tau_m}{(1-u_m x)h_m(x)}.
\end{align}

The integrals above can be, in principle, written in terms of hypergeometric function, but it does not really help to understand their properties. One can show, however, that these integrals are simply related:
\begin{align}
	I_m^1(a,b)- I_m^2(a,b) = h_m^{-1}(a) - h_m^{-1}(b).
\end{align}

The resulting analytic generating function of the mRNA steady state distribution is given by
\begin{align}
	G_m(x)\! =\! h_m(x)\!\left[I_m^1(u_s,x)G_m(u_s)\! +\! I_m^2(u_s,x)G_s(u_m)\right].\label{result-m}
\end{align}

The probability to find an empty system in terms of the reduced distributions can be interpreted as either $\PS_m(0)$ or $\PS_s(0)$ or a combination of both. Due to the symmetry of the representation $\PS_m(0) = \PS_s(0) = \frac{1}{2}\PS(0,0)$, and we obtain from \eq{result-m} 
\begin{align}
	h_m(0)\left[I_m^1(0,u_s)G_m(u_s) + I_m^2(0,u_s)G_s(u_m)\right] = \nonumber\\
	h_s(0)\left[I_s^1(0,u_m)G_s(u_m) + I_s^2(0,u_m)G_m(u_s)\right].\label{cond-0}
\end{align}
Here, the definitions of sRNA's functions $I_s^1(a,b),I_s^2(a,b)$ and $h_s(x)$ are symmetric with respect to the permutation of indices of the corresponding mRNA's functions.

Finally, one derives
\begin{align}
	G_m(1) + G_s(1) = \sum_{m}\PS_m(m)+\sum_{n}\PS_s(n) = 1,\label{norm}
\end{align}
which is the conservation of probability condition.

Combining all conditions together: \eq{analyticity}, the symmetric equation for sRNA, \eq{cond-0}, and \eq{norm}, we obtain a set of four {\em linear} algebraic equations for four unknowns: $G_m(1), G_m(u_s)$ and $G_s(1), G_s(u_m)$.

The generating functions $G_m(x)$ and $G_s(x)$ immediately follow from \eq{result-m} and the symmetric equation for sRNA. These generation functions can be used to study various aspects of sRNA regulation. 

A particularly interesting quantity is the mean regulated level of mRNA. 
Indeed, the mean protein level is linearly proportional to the one for mRNA \cite{ozbudak02,golding05}. The coefficient of the proportionality is the same for regulated and unregulated systems. Hence, one can calculate the repression in {\em protein} expression due to sRNA regulation by comparing regulated and unregulated mRNA levels.

The regulated mRNAs average can be obtain by setting $x=1$ in \eq{steady-m}:
\begin{align}
	\ave{m}_r = \ave{m}\left[1- \frac{\frac{1}{2}I_s^2(u_m,1)}{1+I_m^2(u_s,1)+I_s^2(u_m,1)}\right]\nonumber-\\ 
	\ave{s}\left[\alpha\frac{\frac{1}{2}I_m^2(u_s,1)}{1+I_m^2(u_s,1)+I_s^2(u_m,1)}\right],\label{mean-m}
\end{align}
where $\ave{m} = n_m \tau_m k_m$ and $\ave{s} = n_s \tau_s k_s$ are unregulated RNA levels. Here, $\alpha \equiv \tau_m/\tau_s$ is a ratio of the mRNA and sRNA lifetimes. The symmetric equation is derived for sRNA level $\ave{s}_r$ by permutation of indices $m$ and $s$.

The ratio of the mRNA and sRNA lifetimes satisfy the following condition
\begin{align}
	\alpha = \frac{\ave{m}-\ave{m}_r}{\ave{s}-\ave{s}_r}.\label{alpha}
\end{align}
Interestingly enough, this condition is extremely robust. It holds for finite values of the coupling rate $\gamma$. Moreover, the condition \eq{alpha} is satisfied for any burst distribution and even for the arbitrary arrival time statistics of the RNAs bursts \cite{elgart09a}! This robustness is a result of the symmetry between mRNA and sRNA, since recombination is mutual and involves the same amount of molecules of each  species.

One of the interesting properties of the regulated mean mRNA level, \eq{mean-m}, is the following. Let us define repression level $R$ as a ratio
\begin{align}
	R = \frac{\ave{m}_r}{\ave{m}}.
\end{align}
For the same values of unregulated levels of mRNA and sRNA,
the maximal repression is achieved for a single transcript at a time, $n_m = 1$ for any value of the parameter $n_s$. This is not surprising, since this limit corresponds to the maximal burst frequency of mRNA. Moreover, decreasing the sRNA transcript number increases repression for the same reason. The transcript numbers $n_m=n_s=1$ correspond to the absolute minimum of the ratio $R$.

An important question is how the deterministic (mean-field) prediction compares with the stochastic one.
From the deterministic evolution equation \eq{mf-m} one derives the steady state regulated level in $\strong$ limit
% \begin{align}
% 	m_r = m\left(1 - \alpha \frac{s}{m}\right),\qquad 1 \ge \alpha \frac{s}{m},\label{MF-m}\\
% 	m_r = 0,\qquad 1 < \alpha \frac{s}{m},
% \end{align}
\begin{align}
	m_r = \ave{m} - \alpha \ave{s},\qquad \ave{s} < s_c,\label{MF-m}\\
	m_r = 0,\qquad \ave{s} \ge s_c,
\end{align}
where $m_r$ and $s_r$ are regulated {\em mean-field} levels for each species. Therefore, according to the mean-field prediction above, the mRNA steady state level vanishes once the critical value of the unregulated sRNA level $s_c = m/\alpha$ is reached.

\begin{figure}[t]
\includegraphics[scale=1.2]{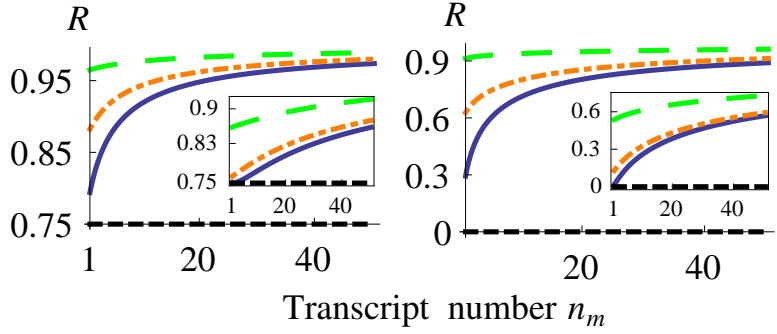}
\caption{The protein repression level $R$ is shown as a function of mRNA transcript number $n_m$. The RNAs' unregulated levels are fixed $\ave{m} = 2\ave{s} = 1$ in both graphs, and the insets correspond to higher unregulated levels $\ave{m} = 2\ave{s} = 10$. The RNAs' lifetime ratio is sampled at values $\alpha = 0.5$ (above critical) and $4$ (below critical) for left to right graphs accordingly. The solid blue curve corresponds to sRNA transcript number $n_s = 1$, dot dashed orange and dashed green curves correspond to $n_s = 5$ and $40$ accordingly. Finally, horizontal dashed line represents the mean-field solution \eq{MF-m}.}\label{fig:MF}
\end{figure}

This mean-field threshold-like solution \cite{levine07,mehta08} is, in fact, a crude lower bound for the stochastic average \eq{mean-m}; see Fig.~\ref{fig:MF}.

The average repression level in protein expression shows significant deviation from the mean-field prediction. This deviation becomes larger with an increase in transcriptional bursting. In particularly, the threshold $s_c$ in protein expression, predicted by the deterministic approach, can be much smaller due to transcriptional noise. This behavior was anticipated by Mehta et al., \cite{mehta08}.

The discrepancy in deterministic description is particularly apparent for small unregulated RNAs' levels. The mean-field solution \eq{MF-m} scales linearly with unregulated RNAs' levels and hence, incorrectly predicts the same repression regardless of RNAs abundance.

The derived generating function can be used to further analyze different aspects of regulation. Quantities, such as noise strength, Fano factor, rare events probabilities, etc., can be derived {\em exactly} from the result \eq{result-m}.

The strong interaction limit may lead to the significant simplification of the spatially extended model models as well. Consider Prey-Predator model on a lattice in this limit (a very similar description holds for the SIR model as well). Any lattice site has now only one residing species at a time.

All local processes (birth and death) occur with transition rates that are linear with respect to the number of particles present at each site of the lattice. The modified diffusion rules can be summarized as follows:

The diffusion of prey from site $i$ to site $j$ always results in an increase in the population of the species residing at site $j$ by one. It does not matter what species was present at site $j$ before the jump.
The diffusion of predators from site $i$ to site $j$ always results in an increase in the population of the {\em predators} on site $j$ by one, with respect to the previous occupancy of any species.

The transition rates due to these modified diffusion rules are linear as well.
The linearity of the all transition rates can be utilized in order to treat models analytically. Therefore, we hope that even for spatially extended models, the strong interaction limit may ultimately lead to an exact solution.

\begin{acknowledgments}
I thank my brother Alex for useful discussions and help with several mathematical aspects of this manuscript. I also thank R. Kulkarni and T. Jia for sharing data from the numerical simulation and for helpful discussions.
\end{acknowledgments}

%\bibliographystyle{apsrev}
%\bibliography{srna}% Produces the bibliography via BibTeX.

\end{document}